# EXTENT: <u>E</u>nabling Appro<u>x</u>imation-Orien<u>t</u>ed <u>E</u>nergy Efficie<u>nt</u> STT-RAM Write Circuit

Saeed Seyedfaraji [1], Javad Talafy [2], Mohammed M. Sabry[3], and Semeen Rehman[1]

[1] Vienna University of Technology (TU-Wien), Vienna, Austria.
[2] Amirkabir University of Technology Tehran Polytechnic (AUT), Tehran, Iran.
[3] Nanyang Technological University (NTU), Singapore, Singapore.

Corresponding author: Saeed Seyedfaraji (e-mail: saeed.seyedfaraji@tuwien.ac.at).

**ABSTRACT** Spin Transfer Torque Random Access Memory (STT-RAM) has garnered interest due to its various characteristics such as non-volatility, low leakage power, high density. Its magnetic properties have a vital role in STT switching operations through thermal effectiveness. A key challenge for STT-RAM in industrial adaptation is the high write energy and latency. In this paper, we overcome this challenge by exploiting the stochastic switching activity of STT-RAM cells and, in tandem, with circuit-level approximation. We enforce the robustness of our technique by analyzing the vulnerability of write operation against radiation-induced soft errors and applying a low-cost improvement. Due to serious reliability challenges in nanometer-scale technology, the robustness of the proposed circuit is also analyzed in the presence of CMOS and magnetic tunnel junction (MTJ) process variation. Compared to the state-of-the-art, we achieve 33.04% and 5.47% lower STT-RAM write energy and latency, respectively, with a 3.7% area overhead, for memory-centric applications.

**INDEX TERMS** Approximation, Magnetic Tunnel Junction (MTJ), Multimedia Application, Spin Transfer Torque Random Access Memory (STT-RAM).

## I. INTRODUCTION

Data movement between the processing unit and main memory is a bottleneck in nowadays system design in terms of energy consumption and performance. These bottlenecks could be aggravated due to high latencies in their memory elements' read/write operation. The above problems became apparent, especially in data-centric applications, e.g., image/video processing, RMS (recognition, mining, and synthesis), computer vision, communication, and networking. However, the work in [1] has demonstrated that such applications can tolerate approximate errors while producing valuable results. Several state-of-the-art modern memory architectures have proposed different methods to alleviate the read/write energy and latencies [2]–[5]. However, to further enhance the read/write energy and latencies, in this work, we exploit the approximate computing concept to design an energy-efficient memory architecture.

Conventional memory technologies, such as SRAM and DRAM, suffer from power issues [6]. Non-volatile memories (NVMs) are among the most promising candidates to overcome the power issues because of various benefits, including the near-zero leakage power in case of power gating while preserving the stored data [7]. Current-controlled magnetic anisotropy (CCMA) is on the NVM technology with a magnetic tunnel junction (MTJ) device that stores one bit by applying bidirectional current. As shown in Fig. 1, spin alignments in two ferromagnetic layers pose resistance of MTJ device mainly determined by current source direction. STT-RAM as a CCMA family member leverages the migration from pure CMOS product to that of MOS/MTJ combination in which the capacity and cost benefits of DRAM, fast performance of SRAM, virtually unlimited endurance in addition to being non-volatile and radiation hard [1] are employed.

The most recent device, Voltage Controlled Magnetic Anisotropy (VCMA), is a new mechanism where the main difference with the STT-RAM is in the manipulation applied to write a new value in the memory device. The former (current-driven MTJ) uses spin-polarized current. In contrast, the latter (voltage-driven MTJ) uses the VCMA effect [8]. The major problem in VCMA is determining the amplitude and width of applied voltage. However, the stochastic behavior of MTJ requires the adaptive voltage pulse to avoid excessive energy consumption [9].

STT-RAM is highly sensitive towards temperature, and even a slight change in the room temperature can significantly impact the rotation process, as studied in [10]. Thermally-assisted switching or thermal-controlled magnetic anisotropy (TCMA) exerts trade-off write time and write energy [11]. Thus, addressing the thermal effect on the switching process can improve write energy and write time. Therefore current-driven switching accompanied by STT-RAM is considered in this paper.

At the cell level, it is observed that STT-MRAM takes 50% less area with 74% reduction in leakage power dissipation as compared to Spin-Orbit Torque (SOT-MRAM). However, the SOT-MRAMs is 4× faster than the STT-MRAM. At the architectural level, SOT-MRAM outperforms STT-MRAM in terms of read/ write energy and latency at the cost of marginal chip-area and leakage-power [12]. In the way of STT-RAM industrial adaptation, some critical challenges should be addressed related to the write energy and write latency [7], [13]–[15]. To successfully conclude a write operation, two possible solutions are: 1) applying a continuous (consistent amplitude) current as long as it is required, or 2) increasing the current amplitude to speed up the write operation [15]. However, both of these solutions incur unnecessary power overhead in the circuit.

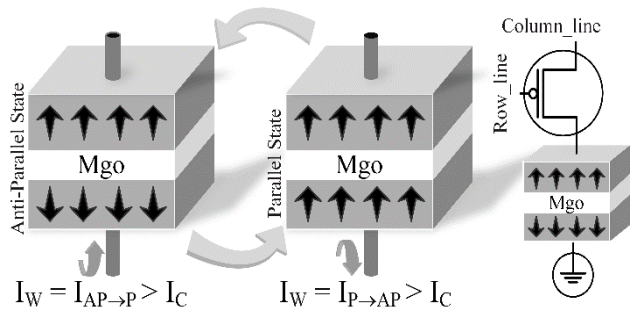

Fig. 1. Change in the state of the MTJ when the current passing through the cell surpasses the critical current ($I_C$) between AP and P states.

There are different state-of-the-art techniques [16]–[21] addressing the write operation of STT-RAM across different levels of system abstractions, i.e., at device, circuit, architecture, and the application level. However, the currently proposed methods ignored the importance of reliability caused by radiation-induced soft errors. Reliable access to memory data, specifically when relaxing, may lower the thermal stability factor and potentially provide conditions for the occurrence of radiation-induced soft error and read disturbance. The striking charged particles can interrupt read operation and intensify inaccuracy in write operation [17], [21].

Systems that deal with a considerable amount of data movement possess an intrinsic error masking capability. They can deal with approximate data that is not visually noticeable by the human eye due to its perception limitation [2], [22], [23]. Therefore, it may not be necessary to keep the accuracy of the processing as high as possible [10]. The works presented in [18], [24] have tried to improve the performance of STT-RAM-based memories exploiting approximation techniques. In [25], a technique has been proposed to achieve different accuracy levels by tuning the MTJ cell's design parameters. They have proposed a hybrid system model including STT-RAM and PCM memory technologies. STT-RAM has been exploited as a scratchpad memory, and the PCM performs the primary memory role. Their approximation knobs, including reading/write pulse width, magnetization orientation, etc., allow them to trade accuracy for latency and lifetime of the memory technology during the design time of the system. The main disadvantage of the proposed technique in [3] is high read disturbance due to the modification of thermal stability factor. The read disturbance refers to a change in the state of the STT-RAM accidentally during the read operation. This problem is due to a shared write and read current path in this memory. In a normal situation, the current amplitude for a read operation is much less (5x–10x) than critical write current amplitude [2]. Nevertheless, even this small current can sometimes lead to a change in the state of the STT-RAM cell. In [18], a quality configurable memory array was designed to store data based on the application demand. The main drawbacks of their study are significant area overhead, long write latency, and high write error rate (WER). Furthermore, imposing a static current will lead to power overhead even though it may not be necessary. The work in [24] proposed a hybrid memory structure as main memory and STT-RAM as a scratchpad memory. Their main goal was to design an approximation evaluation scheme that trades different knobs for the desired accuracy. To summarize, the state-of-the-art has either focused on injecting a static current to the circuit, which leads to power and area overhead, or had approximation over the physical parameters of the MTJ cell, which leads to read disturbance and soft errors induced by radiation particles in harsh environments.

This paper proposes an energy-efficient write circuit by considering four different levels of current injection as an approximation variable. Our technique is based on the thermal behavior of the MTJ cell, whereby tuning the source voltage along with tuning the $V_{th}$ of transistors, we have achieved the optimal power and minimum WER. In a nutshell, the contributions of this paper are: 1) We alleviate the stochastic behavior of the write mechanism in STT-RAM as a spintronic memory device such that write efficiency is accelerated. 2) We proposed a circuit-level memory driver to store the data in diverse approximate modes that the applications with varying error resiliency can use. 3) Enhancing the robustness of our technique by analyzing it under the effects of soft errors and process variations for CMOS and magnetic tunnel junction (MTJ). Compared to the state-of-the-art, we achieve 33.04% and 5.47% lower STT-RAM write energy and latency, respectively, with a 3.7% area overhead, for memory-centric applications (see section IV).

The rest of this paper is organized as follows: section II presents the background of the state-of-the-art methodologies and their reliability challenges. Section III presents our method. The experimental results and evaluation of our method are presented in Section IV, followed by the conclusion in Section VI.

## II. RELIABILITY CHALLENGES IN STT-RAM

STT-RAM is equipped with an MTJ device composed of three layers (pinned/up, thin barrier/middle, free/down) [11]. Read and write operations are carried out by applying currents smaller and larger than the critical switching current ($I_C$) across the MTJ. If the magnetic orientation of these two layers is parallel compared to each other (*P* state), the cell's resistance is in a low-resistance state (i.e., "logic-zero.") On the contrary, if the layers magnetic orientation is in the antiparallel (*AP* state), then it puts the memory element in a high-resistance (i.e., "logic-one") (See Fig. 1) [24].

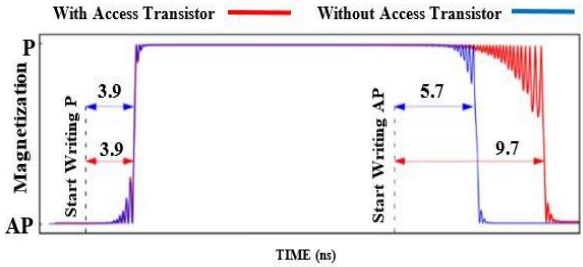

Fig. 2. Stochastic write behavior in the STT-RAM [2]

### A. Write error rate & thermal stability factor

In order to replace SRAM or DRAM in future computing systems, read and write operations must be carried out reliably while considering energy efficiency and delay of accessing the memory element. Current direction crossing through MTJ determines whether the stored data is "1" ("logic-one") or "0" ("logic-zero"). The write operation in STT-RAM is stochastic, which means different currents (density and duration) are required to conclude a successful write operation (see Fig. 2). Moreover, P to AP transition (writing "logic one") takes more time, as shown in Fig. 2, Fig. 3, and Fig. 5. The difference in write current density is due to the current direction passing through MTJ because when electrons flow from the pinned layer to the free layer, they are affected by a strong magnetic field, and they quickly align free layer dipoles. While in the opposite direction, electric particles first enter a weak magnetic field, they need more density or time to reverse the free layer dipoles and write the desired data into MTJ. In other words, the write failure probability in "P" → "AP" transitions is much lower than that of "AP" → "P" transitions [8], [26]–[29]. Increasing write energy compensates for the delay and vice versa. Stochasticity of write operation gets more severe when added to stochastic thermal effects. The thermal stability factor (TSF), also denoted by $\Delta$, of the MTJ device, is an important parameter because it affects the data retention capability. Higher energy barrier of the MTJ device more stabilize the magnetization in current status, resulting in longer retention time. For an in-depth understanding and accurate prediction of the low probability of WER, micromagnetic effects must be considered. Previously WER, calculation considering this effect have been carried out using two cases of 64 and $10^3$ independent stochastic simulations in [30], [31] and [8]. Thus here, WER thermal dependence is defined in the Eq. 1 as follows:

$$WER_{bit}(t_w) = 1 - exp\left(\frac{-\pi^2 \times (I-1) \times \Delta}{4 \times (I \times exp(C \times (I-1) \times t_w) - 1)}\right), I = \frac{I_w}{I_c} \quad (1)$$

$$P_{WER} = 1 - exp \frac{-\frac{\pi^2}{4} \times \left(\frac{I}{I_c} - 1\right)}{\left(\frac{I}{I_c}\right) \times exp\left(\frac{2\alpha\gamma H_k t \times \left(\frac{I}{I_c} - 1\right)}{1+\alpha^2}\right) - 1} \quad (2)$$

$$P_{WER} = e^{\left(\frac{-t_{wr}}{t_{sw}}\right)} \quad (3)$$

Where $I$ is the write current, $I_C$ is the critical current of MTJ cell for write, $I_W$ is the write current, the $t_w$ is the switching time of the MTJ, and the $C$ is a technology-dependent parameter. Stochastic behavior found its root in the process variation that affects both access transistors' threshold voltage and MTJ cell resistance. The process variation will result in the circuit's high resistance and low resistance being non-deterministic, which increases the write error rate due to incomplete writes. The probability of incomplete write operations could be calculated based on Eq. 2 and Eq. 3. $\gamma$ is the gyromagnetic factor, $\alpha$ is Landau-Lifshitz-Gilbert-Damping constant, and $H_k$ is the effective anisotropy field, $t_{wr}$ is writing pulse duration, and $t_{sw}$ is the average switching delay of MTJ cell. In the following, we present a write driver circuit that uses different levels of write current density based on the need of the application to improve performance in STT-RAM.

### B. Random error intensifying write error rate

An important consideration about STT-RAM is taking relaxing into account for reliable, fast, and energy-efficient operations. As shown in Fig. 6, the tunnel magneto-resistance ratio (TMR) decreases as the temperature elevates. TMR is a parameter that is defined based on the $R_{AP}$ (high-resistance) relation to $R_P$ (low-resistance). TMR=($R_{AP}$-$R_P$)/$R_{AP}$. Fig. 7 shows the relation between switching voltage and switching time of MTJ cell under different temperatures. It can be seen that for a specific switching voltage, as the temperature increases, the switching time decreases, and for a specific switching time, a lower voltage is necessary to switch the cell content as the temperature increases. The higher write current results in higher retention time. Meanwhile, there is a risk of oxide barrier break down as a hard error [44], [45]. High write energy is required to improve retention time, and high energy applied to cells reduces endurance. Endurance, the number of times the device can be switched before the failure occurs, is proportional to the ratio of time to failure to switch times. Therefore, retention and endurance are strongly inversely linked. Hence it seems STT-RAM has low endurance and high reliability compared to VCMA since high write energy is required to change the direction of free layer spins. On the other hand, there is no dependence on the magnetic field. If there is, it will affect adjacent cells in the submicron area. Subsequently, reliability is high even though scalability is limited. So, it is necessary to consider the retention/write-speed tradeoff in memory unit design. Write speed is very often a super-exponential function of applied voltage. Therefore, an enhancement in write speed or immense stress results in enhanced retentions time and is harmful to endurance.

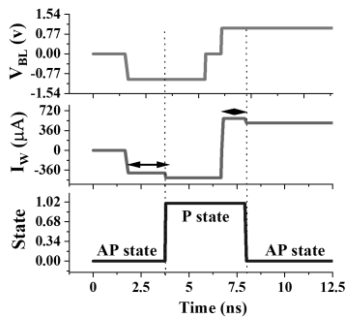

Fig. 3. Write operation of MTJ. Switching between Anti-Parallel (AP) to Parallel (P).

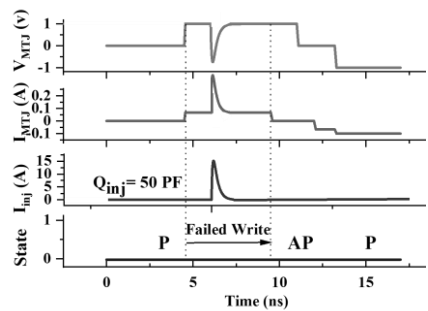

Fig. 4. Failed write operation of MTJ

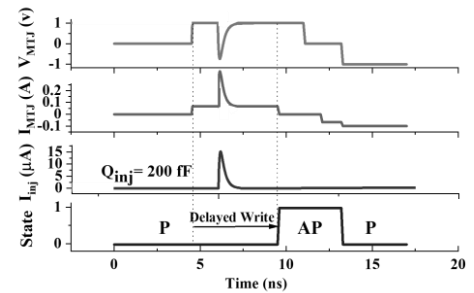

Fig. 5. Delayed write operation of MTJ

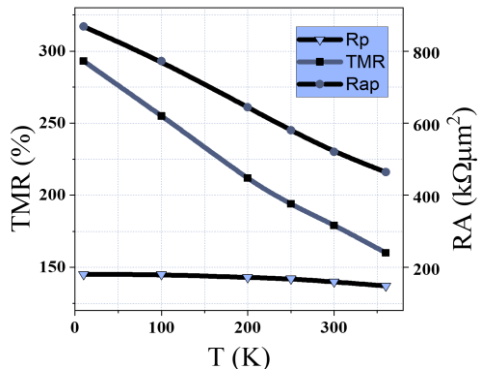

Fig. 6. Tunnel magnetoresistance ratio relativity to temperature [10], [14]

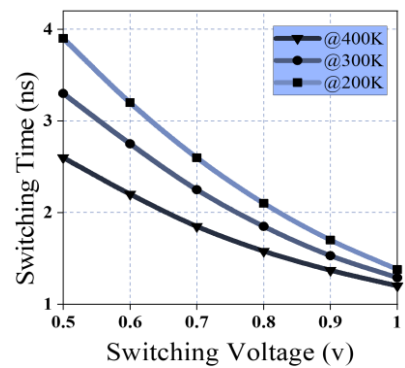

Fig. 7. Switching time voltage dependability to temperature [10], [14]

The state-of-the-art technique, [12], attempts to improve the write performance of STT-RAM memories. However, the main concern was to find a way to decrease the WER as much as possible. Increasing the size of transistors in older technologies, such as 32 nm, improves the reliability against high energy particle strikes, but in the technology below 32 nm, the effect of the particle is not reduced by increasing the size of the transistor. However, the larger the size of the transistor, the channel's resistance decreases, and the memory element's switching speed increases as the current flow rate increases. To investigate the effect of particles on write operation double exponential current source is used. Fig. 3 shows the operation of the write circuit without the occurrence of a soft error.

Fig. 4 and Fig. 5 show the effect of the collision of the charged particle on the degradation of the writing process. Simulation results have shown an injected charge of $Q_{inj}$=50 Pf and $Q_{inj}$=200 Ff, respectively. In the conventional write circuit, there are both pMOS and nMOS transistors, so the current source direction is out of the stroked node (negative glitch) and into the node (positive glitch). Depending on the strength and type of particle, there are different effects. The behavior of the STT-RAM over the thermal fluctuation is expressed in Eq. (4), (5), and (6) [2].

$$I_c = \alpha \left(\frac{\gamma \times e}{\mu_B \times g(T)}\right) \mu_0 M H_k V_k V_{si} = 2\alpha \left(\frac{\gamma \times e}{\mu_B \times g(T)}\right) E \quad (4)$$

$$t^{-1} = \frac{1}{\tau_0 \ln\left(\frac{\pi}{2\theta_0}\right)} \times \left(\frac{1}{\lambda I_c} - 1\right) \quad (5)$$

$$g(T) = \frac{\sqrt{TMR(T,V) \times (TMR(T,V) + 2)}}{2 \times (TMR(T,V) + 1)} \quad (6)$$

In these equations, α is the Gilbert damping coefficient, γ is the gyromagnetic ratio, e is the electron charge, $\mu_B$ is the Bohr magneton. $\mu_0$ is the permeability of free space, MS is the saturation magnetization of MTJ cell layers, $H_K$ is the magnetic anisotropy field, $V_{Sl}$ volume of free layer and E is the energy barrier. $\theta_0$ is the initial angle of the magnetization direction of the free layer, which is thermally distributed and can be calculated by [6]. $\tau_0 \sim 1.0$ ns is the relaxation time, λ is a coefficient λ= 0.2333, and I is the writing current.

### A. Related Works

All the previous approaches can be categorized into two main categories: 1) circuit level designing and 2) architecture level designing [32]. In the first category, there are approaches that trying to tune the physical parameters of MTJ cells [7]. Although the method of [23] has some improvement in terms of energy, it has significant side effects like manufacturing limitations that make its physical realization face considerable design hardships. In paper [23], the authors increase the thermal stability factor (Δ) from 20 to 60. They calculate the mean time to failure (MTTF) for each step. Another technique is based on trying to cut the write current just as the write operation concluded [15], [25]. In this method, the write driver of the circuit attempts to finish the task within a particular time duration (10ns). If the written data and the input data are equal, then the write elimination function is activated otherwise, continuously, the current increases through the write injection circuit to guarantee a successful write operation. The overall write latency remains unacceptable. State-of-the-art methodologies are based on accepting bit errors during a write operation and then using a reliable error correction code (ECC) to proceed with the data [13]. Despite all the benefits of using ECCs, it imposes a significant decoding latency overhead on reading and writing operations. Most of the previous approximation techniques present their ideas based on either tuning the physical parameters like thermal stability [3] or designing a memory architecture based on the desired approximation level [24]. The main strength of our novel idea is that we have exploited the thermal behavior of MTJ cells to design a complete memory system at the circuit level that is both energy aware, and process variation immune. Our approach offers high performance and a decrease in the energy consumption while incurring a bounded area overhead. Write operation can be carried out at different levels of approximation based on the amount of current passing through the memory cell. These levels are usually designed based on the application requirements. In EXTENT, we have developed a novel approximation-oriented technique at the transistor level, resulting in a trade-off between energy, latency, and quality. In this section, we first define some parameters over the approximation in STT-RAM memories. Afterwards, we will describe our circuit-level approximation technique in the write circuitry. Finally, we evaluate our circuit under a comprehensive simulation.

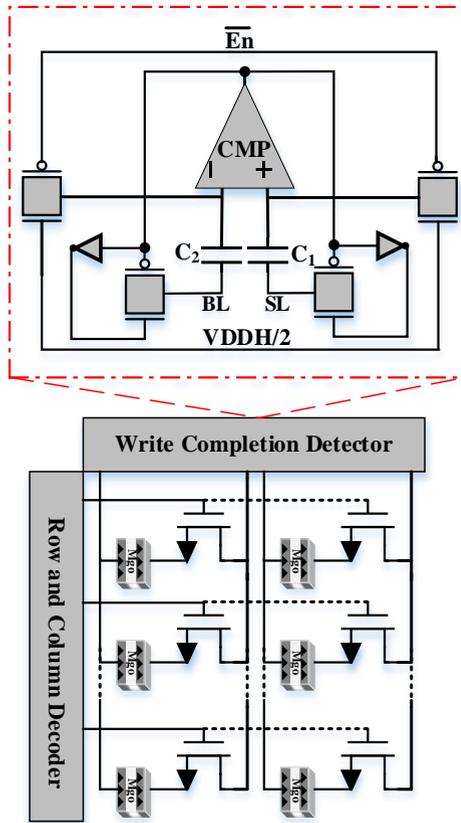

Fig. 8. Our novel EXTENT memory array structure

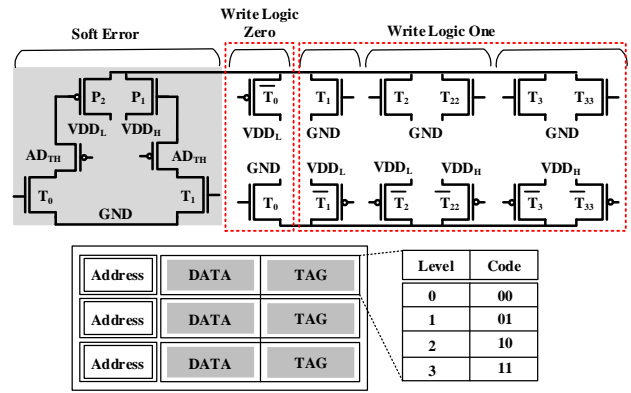

Fig. 9. Our novel Proposed EXTENT write circuit

## III. EXTENT METHODOLOGY

### A. Circuit Level Support

Fig. 8 presents the memory structure of our proposed technique including that includes a row/column decoder, and a write completion detector (WCD). This paper does not contribute to the WCD modules, and it is placed here for clarity. Consider that this module is shared between all memory cells vertically. The write circuit of our proposed technique is presented in Fig. 9. This write driver will be part of a row decoder and consist of a soft error-tolerant and an approximate write current injector. As mentioned before, for writing "logic zero," the access transistors are connected to VDD_Low (*VDDL*), and for the "logic one," there are three available methods to change the state from parallel to anti-parallel state. It is worth mentioning that the concept of dual_VDD will impose a negligible overhead in the fabrication process. There are different approaches to handling voltage domains, including level converter, clustered voltage scaling (CVS), dynamic voltage scaling (DVS), or dynamic voltage/frequency scaling (DVFS) [33], [34]. However, since we only need two levels of voltage in our approach, we could exploit a bandgap circuit to make two voltage levels. A bandgap voltage reference is a temperature-independent voltage reference circuit widely used in integrated circuits. It produces a fixed voltage regardless of power supply variations, temperature changes, or circuit loading from a device [46]. We also modified the threshold voltage of the transistors connected to high voltage (*VDDH*) to attain the incremental heat dissipation because of the leakage current path through the transistors ($\overline{T_{22}}, \overline{T_3}, \overline{T_{33}}, T_{22}, T_3, and\ T_{33}$). The write driver acts, based on the quality decoder's decision, activates one part of the write driver in the memory circuit. If the input data is "logic-zero," then $T_0$ and ($\overline{T_0}$) connected to *VDDL*, will start to inject current on the memory element. On the other hand, if the input data is "logic-one" then there are three possible options for injecting the write current. The quality controller will transfer these final choices of these three options, and there will be tagged based on the importance of the data. In the first category [7], [8], and [9], which the data ("logic-one") has minor importance, the transistor $T_1$ and $\overline{T_1}$ will be responsible for writing the circuit. As can be seen in Fig. 9, these two transistors are connected to *VDDL* so that the amount of current pathing through the cell will be the least, so there will be some write errors on the limited time intervals. In the second scenario, the data's priority is medium, so there will be two pair transistors ($T_2, \overline{T_2}, T_{22}, \overline{T_{22}}$) involved in writing the input data. The threshold current density in the MTJ cell can be expressed as the Eq. (7), (8), and (9).

In these equations, *t* is the width of the free layer of the MTJ cell, $g(\Theta)$ is the efficiency, $H_S$, $H_{ki}$, and $H_{dip}$ are the perpendiculars applied perpendicular anisotropy, and dipole fields from the pinned layer acting on the free layer, respectively. At the architecture level, the incoming data is tagged based on the priority of the data which will be assigned by the programmer via our provided application programming interface from the application level and is categorized into four different levels of priority from 00 to 11.

$$J_{CO}^{P \to AP} = \left(\frac{\alpha \gamma e t M_s}{\mu_B \times g(0)}\right) \times [(H_{ex} + H_{dip}) + (H_{ki} + H_d)] \quad (7)$$

$$J_{CO}^{P \to AP} = \left(\frac{\alpha \gamma e M_s t}{\mu_B \times g(\pi)}\right) \times [(H_{ex} + H_{dip}) + (H_{ki} + H_d)] \quad (8)$$

$$g(\theta) = \frac{P}{(2 \times (1 + P^2)\cos\theta)} \quad (9)$$

Afterwards, EXTENT starts to read/write the data in the circuit. At first, quality control detects the priority of the data and activates the corresponding write enable signal. The address will be fetched from the address decoder, and then if the CMP module grants the write permission, it will be only approved upon new and not repetitive data. Afterwards, the write module injects current. During the operation, the CMP module senses the cell's value and cuts off the write current immediately when the value of the cell changes. CMP is a device that compares two voltages (or currents) and outputs a digital signal indicating which is more prominent. It has two analog input terminals, V+, V-, and one binary digital output Vo. Please note, there is no emphasis on using a specific type of comparator in this manuscript, but a critical parameter that needs to be considered is the resolution and compatibility with the CMOS process. For our evaluation using HSPICE simulation setup, we have considered the circuit design presented in [20]. To find a correlation between the body-biasing and the threshold voltage of the transistor, Eq. 10 can be applied. The sub-threshold current can be calculated by tuning this threshold parameter and using Eq. 11. As the sub-threshold parameter increases, the mobility parameter of transistors also increase, according to Eq. 12.

$$V_{th} = V_{th0} + \gamma \left(\sqrt{|2\Phi_F + V_{SB}|} - \sqrt{|2\Phi_F|}\right) \quad (10)$$

$$I_{DS} = I_s \times 10^{\frac{V_{GS}-V_{th}}{S}} \left(1 - 10^{\frac{V_{GS}-V_{th}}{S}}\right) \quad (11)$$

$$I_{DS} = \mu C_{ox} \left(\frac{W}{L}\right)\left(V_{GS} - V_{th} - \frac{V_{DS}}{2}\right)V_{DS} \quad (12)$$

The final result of an increase in mobility will be an increase in the transistor's temperature (according to Eq. 13). In these equations, $V_{th0}$ is the threshold voltage for the initial state of VSB, and $\Phi_F$ is the surface potential. W, and L are width and length of the transistor. Cox is gate oxide capacitance per unit area. μ is the carrier mobility of the transistor and depends on the temperature indicated by Eq. 11-15.

$$\mu(T) = \mu(T_r)\left(\frac{T}{T_r}\right)^{-k_u} \quad (13)$$

$$P_{SW} = 1 - \exp\left(\frac{-t_p}{\tau}\right) \quad (14)$$

$$\tau = \tau_0 e^{\Delta\left(1 - \frac{V}{V_{co}}\right)} \quad (15)$$

In these equations, $P_{sw}$ is the switching probability of the MTJ cell. $t_p$ is the duration of the voltage pulse period. $\tau_0$ is an initial thermal state of MTJ cell, which is 0 K. Finally, according to Eq. 14 and Eq. 15, we can extract the correlation between temperature and the probability of changing in the free layer. In these Equations, $P_{sw}$ is the switching probability of the MTJ cell. $t_p$ is the duration of voltage pulse period. $\tau_0$ is initial thermal state of MTJ cell which is 0 K. By increasing the size of the transistor, it is not possible to protect it against high energy particle strikes in below 32nm technology nodes. For this reason, in our proposed circuit, we increased the number of parallel transistors without increasing the size of the transistor, such that our designed circuit can be used for any technology with any transistor size and will work appropriately. Depending on the application, two transistors with an adjustable threshold can be fabricated with varying $V_{th}$. In a scenario, consider a particle strike over the N1, and transistors $T_1$ and $T_2$ are in one state. The negative glitch of particle may cause the adjustable transistor switch to on-state; consequently, $P_2$ will switch to on-state, and the unexpected glitch will be compensated. The overview of the complete memory structure has been presented in Fig. 8.

### B. Software Support

In order to execute an application on our EXTENT architecture, there is a need for a software level interface. By exploiting the interface, first, we need to identify which part of the program could be approximated and which will tolerate a certain amount of inaccuracy. Any inaccuracy in the application's flow control could not be tolerated since it will lead to the failure of the whole system. The concept of identifying the priority of data has been studied before, where the authors do the partitioning of the data and tag them with high or low priority to improve DRAM energy consumption [21]. To make the EXTENT framework as efficient as possible, we integrated the available practitioners such as Rely [35] and Accept [36] to our system. Please note that, this paper presents hardware level contribution, and future effort will be devoted to designing and implementing an intelligent partitioning approach. For the interest of this paper, we will present the implementation of application programming interface library that maps the priority tags to a standard hardware command. Our interface provides two simple tags, including "high_prioriy" and "low_prority," that takes virtual address, word size, and priority factor as input and send the command as an output.

```
#include <additional Libs>
input *Img
integer priority_level = default;  //High:11, Low:10, Normal:00
float physical_variables; //Humidity, Temprature, UV

integer Image_load(Graphics::TBitmap *bmp, AnsiString name, integer *_alpha);

//********************Processing Block********************//
loop all_pixels in Img :
            avg = (Pixel.Red + Pixel.Green+Pixel.Blue)/3;
            Pixel = avg();
            priority_level = "10"

     if (procedure = start)
     {
            physical_variables = get physical_variables();
                     if (condition_first && condition_second)
                               priority_level = "10";
                     else if (condition_second && condition_third)
                               priority_level = "11";
                     else
                               priority_level = "00"
     }
```

Fig. 10. A pseudo-code exemplifying the compiler operation

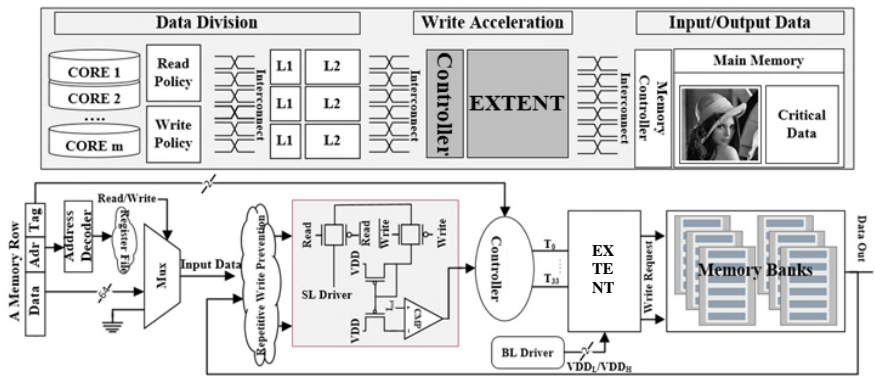
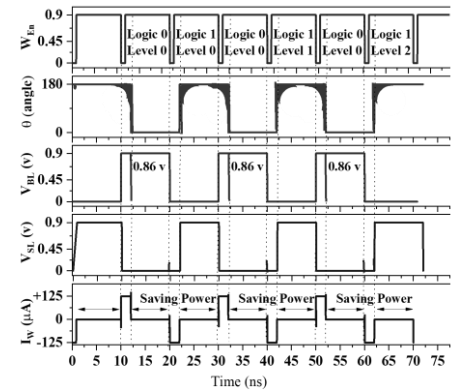

Fig. 11. The proposed architecture exhibiting the top-down perspective of proposed method.

Fig. 12. Simulation waveform of different levels

We consider a multicore processor with a dedicated L1 cache and a shared last-level cache (LLC). Specifically, we integrate our proposed architecture inside LLC since it contains most of the data with a different set of priorities. Our STT-MRAM based LLC will bring noticeable improvement to overall power consumptions for a different set of applications. Our proposed circuit could be implemented in a different architecture setting, including single core or multicore. We will deal with write operations with different priorities in the following steps. First, the priority level of the incoming data will be extracted by the interface mentioned above, and then at the hardware level, the priority aware module will activate the corresponding write driver to execute the operation. The pseudo-code presented in Fig. 10 is a possible scenario in which a change in a physical parameter of the experimental environment could be transferred from application via our interface to the proposed circuit.

### C. Architecture Support

It can be seen in Fig. 13 that the percentage of transition going from "logic-zero" to "logic-one" is high enough in the L2 cache to consider them as a substantial potential to alleviate the approximate computing in these benchmarks [37], [38]. In both cases, on average, 80% of all cache access patterns includes high energy-consuming during the transition of "logic-zero" to "logic-one" [21], [37]. Most of the data in the L1 cache belongs to the group of data flow, and cannot be approximated to achieve energy efficiency (i.e., loop counters, temporary variables, etc.).

Our novel EXTENT circuit will be part of the L2 cache, which has the most potential for tolerating approximation without causing any permanent failure in the system due to high refresh rate of the data. The proposed architecture for the EXTENT is shown in Fig. 11. The input data from the application level will be passed to the EXTENT controller via the provided API. Furthermore, we modify the system architecture to pass the quality parameters to the write driver of our circuit and simultaneously update the EXTENT table. The proposed architecture is optimized to achieve the most performance while conveying the quality value submitted from the upper layer. EXTENT will reserve the reported quality for each memory block to decrease the latency of multiple access to the same block. The application sends a "priority_level" command via the API and the address to the EXTENT controller. At first, the controller updates the EXTENT table with the incoming data and then transmits the physical address along with the denoted quality to the write driver circuit. When the application asks for memory access, first, the EXTENT table will be searched, and if there is a hit, the quality controller will transmit the value. Otherwise, the writer's default value will be considered the proposed quality.

### IV. Cross-Layer Evaluation:

#### A. Architectural Evaluation

This section depicts the architecture evaluation of the proposed architecture for different industrial benchmarks (see Fig. 14). We exploited and rectified the cache inside the GEM5 full-system simulator to implement the proposed method for the homogenous architecture [39]. Accordingly, we applied our EXTENT architecture to the LLC. We modeled the quality decoder similar to a random fault injection scheme during the write operation for "logic-one" in both level-1 and level-2. The fault injection follows the WER probability function mentioned in Eq. 3 and Eq. 4 for each memory element following a uniform distribution. Table 2 depicts the detailed information regarding the simulation platform of our method. System-level accuracy would either follow the circuit's write error rate (WER) or improve (due to application-level masking effects). As all examined techniques are mapped to the same workloads, application-level error masking effects would be experienced similarly. Hence, we can assess system-level accuracy using WER. Our approach showcases either a similar or better write error rate to the state-of-the-art. The normalized average energy usage reported in Fig. 14 would not be affected considering this assumption. In fact, we may report better energy benefits if we relax our accuracy to match the values reported in previous works.

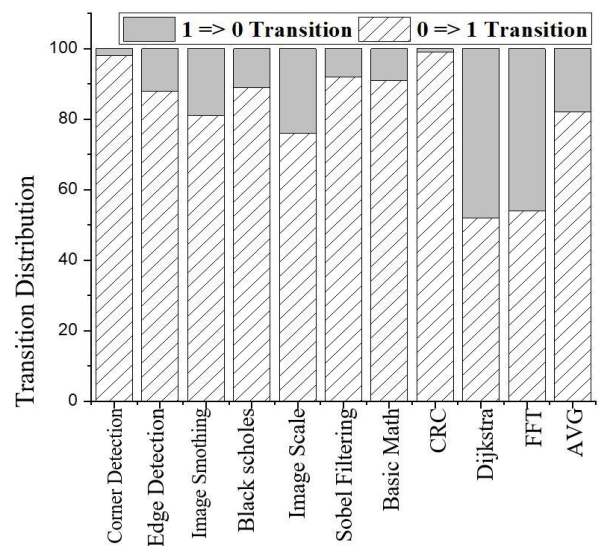

Fig. 13. Access Pattern for different workloads in L2 Cache [30][34].

Table 2. Details of Caches configuration

| | |
|---|---|
| CPU | Quad-core, 1GH, out-of-order |
| L1 Cache (Ins. And Data) | 32KB, 4-way set associative, 64B B-size, Nonblocking, private |
| L2 Cache | 1MB, 8-way set associative, 64B B-size, shared, TASTE active. |

Table 1. EXTENT VS. STATE-OF-THE-ARTS

| | Basic Cell | [18] | [21] | This Work | [40] |
|---|---|---|---|---|---|
| **Area [mm$^2$]** | 1.31 | 1.37 | 1.31 | 1.46 | 1.41 |
| **Write Latency [ns]** | 19.0 | 2.2 | 7.3 | 6.9 | 7.8 |
| **Write Energy [pJ]** | 1046.0 | 503.6 | 393.3 | 337.2 | 356.9 |
| **Self Termination** | No | No | No | Yes | Yes |
| **Monitoring** | None | Continuous | None | Continuous | Continuous |

### B. Circuit Evaluation

For calculating the exact value of Vth-nmos and Vth-pmos, simulation is done by considering the circuit under 400K. Considering the relation in the Eq. 11-15, these temperatures lead to a decrease in the overall write latency of the circuit. The exact value for VDDL is equal to 0.86001v and calculated by considering the VDDH of the circuit, total power consumption, and latency overwriting "logic-zero" as optimization parameters [11]. The EXTENT has been simulated under various process corners. We confirmed the immunity of our circuit towards process variation effects (see Fig. 15 and Fig. 16). The simulation waveforms of EXTENT shown in Fig. 12 proved that this circuit can identify the input data. If the new input data is equal to the stored data in the cell, the system immediately cuts the current to prevent any repetitive write operation, which leads to further energy reduction. The parameter Θ represents the angle between the free and fixed layers of MTJ cells. The minor disturbance seen on the Θ waveform happens basically because of initial current passing through the MTJ cell before the system identifies it as a repetitive write attempt. In those write operations, where the input data is not equal to the stored data, it can be seen that the parameter Θ has a range between 0, and 180, respectively, to change the magnetization orientation of the free layer and fix layer. In a non-repetitive write based operation, after the write operation concludes, the resistance of the cell will change. It is worth mentioning that the write enable signal pulse width has been decided to be equal to 10ns which is equal to state-of-the-art approaches [2][17][20][37]. These changes in resistance of the cell will cause a minor disturbance over either VBL or VSL. This disturb will be used in the comparator module to cut the write current right after the operation is concluded. One of the writing drivers will be activated to increase or decrease the write error probability on the importance of the input data (there is four-level of approximation in the circuit). Writing "logic-one" usually takes more time and energy than "logic-zero" (2.5X). TABLE 1 shows a comparison of the essential parameters of STT-RAM memories. EXTENT circuit improves the main parameters of STT-RAM with a negligible area overhead compared with state-of-the-art solutions. In all cases, "Read-Energy" and "Read-Latency" parameters are equal as there is no change in the read circuit of all these circuits. For the comparison result, it is worth mentioning that the idea of the reported paper [21] is based on discussing the effect of the thermal stability factor on the image's output quality. To make a meaningful comparison, we first implemented the approach presented in [25] with a considered 32nm technology node and increased the thermal stability factor from 10 to 70 using SPICE simulation to get the write latency and write energy. For that, we needed circuit-level evaluation. This step is reported in TABLE 1. Afterwards, we pass this result into our Gem5 full system simulator to extract that paper's output result for our applications. The results have been extracted for [18] and [21], and an average has been taken for a range of their reported results. The circuit-level characteristic has been extracted via SPICE simulation for all available approximation levels in our paper, including fully approximate, fully accurate, and partially approximate. Then system-level evaluation has been made on the same Gem5 machine for result comparison. We have performed a circuit-level simulation exploiting HSPICE for a 3-level write driver circuit. The STT-RAM model compatible with SPICE in all these simulations is presented in [41], and the 32nm PTM CMOS technology model is presented in [42]. Table 3 summarizes the basic configuration of the STT-RAM model. In all the analyses, the supply voltage is set to 0.9 (V) and the pulse width duration equals 10ns. This technique is generalized and can be applied to any other model available.

Table 3. MTJ CELL PHYSICAL PARAMETERS

| Parameter | Description | Default Value |
|---|---|---|
| **Area** | MTJ Cell Surface | 16e$^{-9}$ (mm$^2$) |
| **TMR$_{(0)}$** | TMR with zero V$_{bias}$ | 200% |
| **T$_{ox}$** | Oxide Barrier Height | 8.5e$^{-10}$ |
| **R.A** | Resistance×Area | 5 (Ω. mm$^2$) |
| **I$_C$** | Critical Current | 200 μA |
| **T** | Room Temprature | 300 °K |
| **Tsl** | Height of the free layer | 1.3e$^{-9}$ |
| **R$_P$** | Low Resistance | 4.2 KΩ |
| **R$_{AP}$** | High Resistance | 6.6 KΩ |

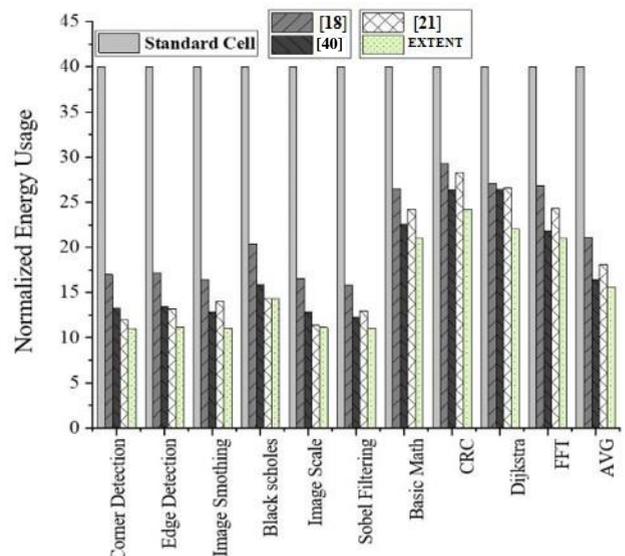

Fig. 14. Normalized energy improvement compared to state of the art

### C. Application Evaluation.

We did profiling over the benchmarks stated in Fig. 13 to obtain their access patern to the L1 and L2 Caches. We consider several applications from the MiBench benchmark suites [35] The amount of energy used for each benchmark can be seen in Fig. 13. It can be seen in Fig. 14 that more energy saving is achieved for all the benchmarks with the majority of "logic_zero" to "logic_one" transition. However, it is worth mentioning that, EXTENT is highly independent of the running application. The MiBench benchmark here has been nominated to compare to the state-of-the-art methodologies. Although the benchmark mentioned above may not fully utilize the CACHE, it will be under heavy workload when the running application is on a closed loop.

### D. Variation Analysis

Magnetization switching agitated by thermal assistance in which current density can be much smaller than the critical density is an energy reduction approach. Due to serious reliability challenges in nanometer-scale technology, quantitatively, reliability and energy consumption simulation under the joint contribution of process, supply voltage, and temperature variations have been studied. Fig. 15 shows the impact of process variation on the minimum writes energy of EXTENT with different sizes of transistors, with and without the consideration of approximation. Thermal activation switching occurs in more than about pulse width duration 10 ns. Variation in the input supply voltage leads to a variation in write current and write performance. Fig. 16 shows the effect of voltage variation on write performance. Process variation affects all the magnetic cells simultaneously. It means the resistance of some memory elements increases, and some will see a decrease in their resistance. Here we considered a standard deviation and mean average for various circuit parameters. Due to the low TMR of STT-MRAM in comparison with other non-volatile memory technologies, process variation improvement would be possible if we can turn down the resistance of MTJs in a parallel state and turn up the resistance of anti-parallel state. Parallelization of two low resistance and serialization of two high resistance MTJs can be an efficient physical level solutions as mentioned in [42].

By considering the voltage spectrum, EXTENT has a range between 3% to 20%, that exhibits no increase/decrease, caused by voltage has been applied to the write current compared with [21] and [40]. In Fig. 15, different technology nodes write energy of EXTENT individually in the presence of software assistance (approximated), and uniform write (completed) is depicted. Effect of variation for approximated write is low (leis between min and max values). On the other hand, write energy for completed write changes between 400 to 1200 Pj; however, approximated write ranges between nearly 0 to 500 Pj. For a fair and equitable comparison with other approaches such as [43], the range of 1, to 10 have been considered ~up to 10%. To study the process variation effect from one side, we considered oxide barrier thickness (10%), FM layer thickness (10%), and resistance of the cell (5%) for a compact PMA STT-MTJ SPICE model [41]. On the other side, we considered 3σ variation over the channel length, width, and the threshold voltage of a commercial CMOS 32nm technology. The process variation also includes the thermal fluctuation on the whole circuit. *Monte Carlo* simulation has been performed 1000 times to obtain a variation of write energy in the presence of process variation in our proposed circuit. Therefore, to evaluate the effects of process variation on the regular operation of our EXTENT, we selected a value as the percentage of variation on mentioned parameters. The mentioned parameters are varied up to 10% from their original values using a gaussian distribution with a standard deviation of 3%.

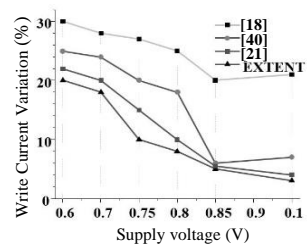
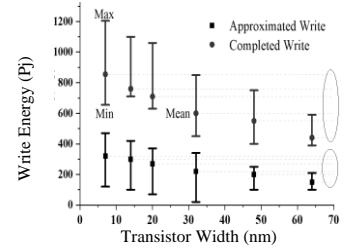

Fig. 15. write current variation   Fig. 16. The effect of transistor width variation on the write energy

## V. Conclusion

STT-RAM leverages the migration from pure CMOS product to that of MOS/MTJ combination in which the capacity and cost benefits of DRAM, fast performance of SRAM, virtually unlimited endurance, and being non-volatile and radiation hard are employed. Suffering from high write energy is the main obstacle of NVMs for replacing the current technologies such as SRAM. In this paper, we have designed a new approximation-based method for improving the write performance of an STT-RAM. Our proposed EXTENT is well-designed according to the need of those applications, which can handle the bounded error in their calculation. First, we proposed our circuit-level write driver scheme based on the approximation technique. Afterward, we have evaluated our circuit at the architecture level. With our proposed technique, we obtained 33.04% and 5.47% gain in terms of write energy and latency, respectively, with a negligible 3.7% area overhead compared to the state-of-the-art.

## ACKNOWLEDGMENT

The authors acknowledge TU Wien Bibliothek for financial support through its Open Access Funding Programme.

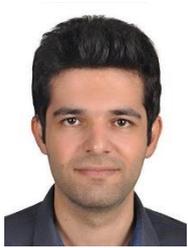

**SAEED SEYEDFARAJI** is currently a University Assistant at Vienna University of Technology (TU Wien), Austria pursuing a Ph.D. degree in Computer Engineering. His research interests include Emerging Non-Volatile Memory Technologies, In-Memory Processing, Bringing Intelligence into Hardware, and System on Chip. He won the Design Automation Conference 2020 Young Fellow (DAC YF 2020) prize along with the Best Team of DAC YF 2020. He took his M.Sc. from Amirkabir University of Technology (Tehran-Polytechnique), Tehran, Iran, and his B.Sc. from Isfahan University of Technology, Isfahan Iran.

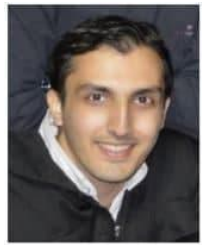

**Javad Talafi Daryani** received her M.Sc degree in computer science and engineering from Amirkabir University of Technology (Tehran Polytechnic), where he spent 3 years in the Design and Analysis of Dependable Systems (DADS) laboratory as a research assistant. His current research interests include both theoretical and experimental aspects of fault tolerant system design, dependable system analysis for future VLSI circuits and systems particularly for computing architectures and memories based on emerging topics like spin-based and resistive-based devices.

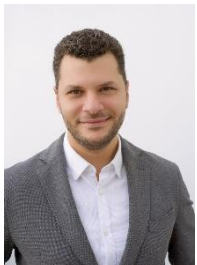

**Mohamed M. Sabry Aly.** Mohamed M. Sabry Aly is an Assistant Professor at Nanyang Technological University, Singapore, and an adjunct research scientist at I2R, A*STAR. He received his Ph.D. degree in electrical and computer engineering from École Polytechnique Fédérale de Lausanne (EPFL), in 2013. He was a postdoctoral research fellow at Stanford University from 2014 till 2017. His current research interests include system-level design and optimization of computing systems enabled by emerging technologies, with particular emphasis on computing systems for artificial intelligence. He is an active close collaborator with top industrial and academic partners such as, Stanford University and TSMC. Prof. Aly is a Senior IEEE member and he was the recipient of the Swiss National Science Foundation Early Post-Doctoral Mobility Fellowship in 2013.

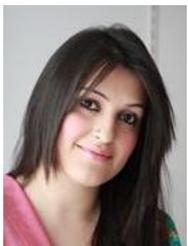

**Semeen Rehman** is currently with the Technische Universität Wien (TU Wien), Faculty of Electrical Engineering, as a tenure-track Assistant Professor. In October 2020, she received her habilitation in the area of Embedded Systems from the electrical engineering faculty, Technische Universität Wien (TU Wien). Before that, she was a Postdoctoral Researcher with the Technische Universität Dresden (TU Dresden) and Karlsruhe Institute of Technology (KIT), Germany, since 2015. In July 2015, she received her Ph.D. from Karlsruhe Institute of Technology (KIT), Germany. She has coauthored one book, multiple book chapters, and more than 50 publications in premier journals and conferences. Her main research interests include dependable systems, cross-layer design for error resiliency with a focus on run-time adaptations, emerging computing paradigms, such as approximate computing, hardware security, energy-efficient computing, embedded systems, MPSoCs, Internet of Things and Cyber Physical Systems. She has received the CODES+ISSS 2011 and 2015 Best Paper Awards, DATE 2017 Best Paper Award Nomination, several HiPEAC Paper Awards, Richard Newton Young Student Fellow Award at DAC 2015, and Research Student Award at KIT, in 2012. She has served as the TPC track chair for the ISVLSI 2021 and 2020 conference and on the TPC of multiple premier conferences on design automation and embedded systems (such as DAC, DATE CASES, ASPDAC). She has (co-)chaired sessions at the DATE 2019, 2018, and 2017 conferences.